\documentclass[aps,reprint,twocolumn,  amsmath,amssymb,superscriptaddress, longbibliography]{revtex4-2}

\usepackage{graphicx,amsmath,amsfonts,amssymb, url}
\usepackage[T1]{fontenc}

\usepackage{graphicx}
\usepackage{dcolumn}
\usepackage{bm}
\usepackage{mathtools}
\usepackage{amsmath, amssymb}
\usepackage{bbold}
\usepackage{url}
\usepackage{orcidlink}
\usepackage{xspace}
\usepackage{hyperref}

\hypersetup{
    breaklinks = true,
    colorlinks=true,
    linkcolor=blue,
    filecolor=magenta,      
    urlcolor=blue,
    citecolor=red
}

\DeclarePairedDelimiter\bra{\langle}{\rvert}
\DeclarePairedDelimiter\ket{\lvert}{\rangle}
\DeclarePairedDelimiterX\braket[2]{\langle}{\rangle}{#1 \delimsize\vert #2}

\newcommand{\red}[1]{#1}

\renewcommand{\vec}[1]{{\mathbf #1}}

\renewcommand{\ket}[1]{\ensuremath{|#1\rangle}\xspace}
\renewcommand{\bra}[1]{\ensuremath{\langle #1|}\xspace}
\newcommand{\ps}[2]{\ensuremath{\langle #1|#2\rangle}\xspace}

\newcommand{\be}{\begin{equation}}
\newcommand{\ee}{\end{equation}}
\newcommand*{\ov}[1]{\overline{\mbox{$#1$\raisebox{0.28cm}{}}}}

\begin{document}

 \preprint{APS/123-QED}

    \title{
    Characterization and Exploitation of the Rotational Memory Effect in Multimode Fibers
    }

\author{Rodrigo Gutiérrez-Cuevas~\orcidlink{0000-0002-3451-6684}}
\affiliation{Institut Langevin, ESPCI Paris, PSL University, CNRS, France}
\author{Arthur Goetschy~\orcidlink{0000-0002-2307-5422}}
\affiliation{Institut Langevin, ESPCI Paris, PSL University, CNRS, France}
\author{Yaron Bromberg~\orcidlink{0000-0003-2565-7394}}
\affiliation{Racah Institute of Physics, The Hebrew University of Jerusalem, Israel}
\author{Guy Pelc}
\affiliation{Racah Institute of Physics, The Hebrew University of Jerusalem, Israel}
\author{Esben Ravn Andresen~\orcidlink{0000-0002-7522-6165}}
\affiliation{Univ. Lille, CNRS, UMR 8523 – PhLAM –Physique des Lasers, Atomes et Molécules, F-59000 Lille, France }
\author{Laurent Bigot~\orcidlink{0000-0002-3541-7039}}
\affiliation{Univ. Lille, CNRS, UMR 8523 – PhLAM –Physique des Lasers, Atomes et Molécules, F-59000 Lille, France }
\author{Yves Quiquempois~\orcidlink{0000-0002-9674-9459}}
\affiliation{Univ. Lille, CNRS, UMR 8523 – PhLAM –Physique des Lasers, Atomes et Molécules, F-59000 Lille, France }
\author{Maroun Bsaibes~\orcidlink{0000-0002-6838-5467}}
\affiliation{Univ. Lille, CNRS, UMR 8523 – PhLAM –Physique des Lasers, Atomes et Molécules, F-59000 Lille, France }
\author{Pierre Sillard}
\affiliation{Prysmian Group, Parc des Industries Artois Flandres, Haisnes Cedex, France}
\author{Marianne Bigot}
\affiliation{Prysmian Group, Parc des Industries Artois Flandres, Haisnes Cedex, France}
\author{Ori Katz~\orcidlink{0000-0002-7746-6349}}
\affiliation{Racah Institute of Physics, The Hebrew University of Jerusalem, Israel}
\author{Julien de Rosny~\orcidlink{0000-0001-8209-532X}}
\affiliation{Institut Langevin, ESPCI Paris, PSL University, CNRS, France}
\author{Sébastien M. Popoff~\orcidlink{0000-0002-7199-9814}}
\email{Corresponding author: sebastien.popoff@espci.psl.eu}
\affiliation{Institut Langevin, ESPCI Paris, PSL University, CNRS, France}

\date{\today}

\begin{abstract}

In an ideal perfectly straight multimode fiber with a circular-core,
the symmetry ensures that rotating the input wavefront leads to 
a corresponding rotation of the output wavefront. 
This invariant property, 
known as the rotational memory effect (RME), 
remains independent of the typically unknown output profile.
The RME thus offers significant potential for imaging and telecommunication applications. 
However, in real-life fibers, 
this effect is degraded by intrinsic imperfections 
and external perturbations, 
and is challenging to observe 
because of its acute sensitivity to misalignments and aberrations in the optical setup.
\red{
Thanks to processing involving a spatial light modulator, 
we efficiently overcome these measurement biases, 
allowing for precise quantification of the RME.
}
We establish an experimental and theoretical framework 
for studying and manipulating the RME in multimode fibers. 
\red{Theoretical predictions are} 
consistent with experimental data and simulations, 
connecting the shape of the angular-dependent correlation of the RME 
to the geometrical properties of the core deformation. 
\red{
This work opens the road for accurate characterization 
of the distributed disorder 
originating from the fabrication process
and calibration-less imaging in multimode fibers.
}
\end{abstract}

\maketitle

\section{\label{sec:intro}Introduction}

Optical fibers present a unique opportunity for minimally invasive imaging deep within the human body.
Most flexible medical endoscopes
utilize multi-core fibers or fiber bundles~\cite{Boese2022endoscopic}. 
Comparatively, multimode fibers (MMFs) offer orders of magnitude higher information density,
allowing, in theory, an increase in image resolution or a decrease in the endoscope footprint~\cite{Turtaev2015multimode}.
However, dispersion distorts the input image, a phenomenon that is exacerbated by mode coupling
introduced by defects or deformations within the fiber. For this reason, 
image reconstruction techniques through multimode fibers hinge on estimating~\cite{Ploschner2015seing} or
measuring the transmission matrix (TM)~\cite{Cizmar2012exploiting,choi2012scanner,papadopoulos2012focusing, papadopoulos2013high-resolution, Jakl2022endoscopic},
i.e., the relationship between the input and output fields of the optical system. Unfortunately, this TM approach is prone to real-time changes due to  dynamic fiber bending 
and temperature fluctuations~\cite{Yammine2019timeDependence}, which
prevent the direct use of a previously calibrated system.

\red{
Similar challenges are encountered in utilizing Multimode Fibers (MMFs) 
for telecommunications, 
where they hold the potential to significantly boost data rates 
compared to their single-mode counterparts. 
Through mode-division multiplexing, 
it is in principle possible to utilize different fiber modes as independent channels, 
effectively multiplying data rates by the number of modes employed,
without substantially increasing cost or footprint. 
However, the occurrence of mode coupling, even in fibers with a limited number of modes, 
i.e. MMFs with fewer than 10 modes, 
currently hinders their application in long-haul communications. 
Enhancing fiber design necessitates a deep understanding of the disorder-induced effects that lead to crosstalk, 
including defects arising from the fiber drawing process. 
Addressing these challenges remains difficult, 
and numerical models frequently used in fiber design often overlook these crucial factors~\cite{Maruyama2017relationship}.
}

\red{
Various fabrication techniques are employed 
based on the fiber type and manufacturer, 
including 
modified chemical vapor deposition (MCVD), 
vapor axial deposition (VAD), outside vapor deposition (OVD), 
and plasma-activated chemical vapor deposition (PCVD). 
One well-recognized challenge across these techniques, 
attributed to their inherent limitations in precision, 
lies in achieving a highly accurate radial index profile, 
particularly when dealing with sharp index changes.
Imperfections in the fiber lead to a mixing of the information carried by the input field. 
This effect is analogous to the randomization of information 
when attempting to recover an image through a scattering medium~\cite{Yu2015recent, Yoon2022recent}. 
To circumvent the necessity of measuring the transmission matrix (TM), 
which implies an invasive procedure, 
an elegant solution for imaging is to exploit the invariant properties of the medium, 
specifically, the angular memory effect~\cite{bertolotti2012non-invasive, katz2014non-invasive}.
}
For a given illumination, even though the output random pattern remains unknown,
the angular memory effect facilitates the shifting of this output speckle pattern in two directions
with minimal to no deformation.
Tilting the input wavefront then allows scanning the object plane with the unknown pattern. 
Recording the reflected or fluorescent signal 
provides sufficient information to reconstruct the image of the hidden object~\cite{bertolotti2012non-invasive}.
While the range of such an effect is constrained, strategies have been proposed to recover images of objects beyond this limitation~\cite{yeminy2021Guidestar},
making the memory effect highly attractive for non-invasive imaging applications.

Building on decades of research in scattering media, 
recent interest has surged in the study of coherent effects in disordered fibers~\cite{bromberg2016control, Devaud2021chromatoAxial}. 
Specifically, the TM approach~\cite{Carpenter2014_110,Carpenter2015Wigner,xiong2016principal,Chiarawongse2018statistical} and random matrix theory~\cite{Yaxin2019coherent,Yaxin2022enforcing} have emerged as particularly useful frameworks for these investigations.
In particular, a close analogous effect 
to the angular memory effect in scattering media 
is observed in the special case of square-core fibers, 
where a translation of the input wavefront results in
a corresponding translation at the output~\cite{CaravacaAguirre2021optical}, albeit with the noticeable presence of artifacts,
which can nonetheless be exploited to recover images~\cite{Mezil2023Imaging, Bouchet2023speckleCorrelation}.
In more typical cylindrical-core fibers,
a similar phenomenon, known as the rotational memory effect (RME), 
has been recently identified~\cite{amitonova2015rotational, Li2021memory}.
This effect is characterized by the rotation of an input wavefront along the optical axis of the fiber
leading to a corresponding rotation of the output pattern, 
even though the latter is unknown.
In principle, this effect could be harnessed for imaging through a multimode fiber for which the TM has not been previously measured.

Nevertheless, since its initial observation, 
no prediction or quantitative description of the RME has been presented. Neither the angular range covered by the RME, nor its dependence on disorder, nor its potential robustness and modularity have been studied or elucidated. 
Furthermore, the manifestation of an angular revival effect, 
leading to secondary peaks in the correlation of the output pattern
at the rotation angle $\pi$, has been observed but also remains unexplained.
An important consideration is that the measurement of the RME is complicated by its high sensitivity to
misalignments and aberrations in the optical system used to inject light into the fiber~\cite{amitonova2015rotational}.
However, these adverse effects can be understood and compensated
using a framework that some of us have introduced in Ref.~\cite{matthes2021learning}.
The procedure involves learning the input and output aberrations by optimizing
a model-based numerical model.
This approach enables the retrieval of an accurate TM of the system, 
even when using imperfect measurements.
Additionally, it provides the transformation needed to physically compensate for the input aberrations,
which can be directly implemented using a spatial light modulator (SLM). 
The numerical compensation for aberrations is a crucial step, 
as it enables precise observation of the RME, 
which would otherwise be rapidly obscured by aberration effects.

In the present article,
we first demonstrate that it possible to measure the RME in MMFs with high accuracy.
\red{
We then introduce a theoretical framework based on precise disorder modeling, which yields analytical predictions for the shape of the RME correlation function, 
supported by numerical simulations of the microscopic wave equation in MMFs. 
Specifically, this framework properly predicts the angular range of the RME as function of disorder, and elucidates the origin of 
secondary peaks in the correlation function. These are attributed to the subtle interplay between mode symmetry and disturbance symmetry.
}
The theory and simulations faithfully reproduce our experimental results obtained on various commercial MMFs. 
\red{
Finally, based on this analysis, 
we propose a method to identify input wavefronts 
that are only weakly sensitive 
to the symmetry breaking introduced by perturbations, 
opening the door to exploiting the RME 
for calibration-less imaging applications.
}

\section{Measuring the RME}

A memory effect is defined in relation to a field transformation.
A perfect memory effect exists when the application of this transformation
before or after  propagation through a given optical system produces the same effect.
For RME to occur,
the rotation operator $\mathbf{R(\theta)}$ must commute
with the optical system's transmission matrix $\mathbf{T}$ of the fiber~\cite{Li2021memory}. 
In this study, we consider only one polarization of the field. The matrix $\mathbf{T}$ therefore links the input field in a specific circular polarization channel to the output field in the same polarization channel.

\subsection{Experimental setup and measurement procedure}

\red{
In principle, the measurement of the memory effect is straightforward, as 
it simply requires rotating the input wavefront and measuring the corresponding output pattern. 
However, any factor that disrupts the rotational symmetry 
of the system leads to a degradation of the measurement. 
In particular, minute misalignments relative to the fiber axis 
and aberrations can alter the observed results~\cite{amitonova2015rotational, Ploschner2015seing} .
This complicates quantitative characterization 
by making it impossible to separate the effects of fiber defects 
from those introduced by the optical setup used for measurement. 
We address this issue by first numerically estimating 
these detrimental effects, 
and then experimentally compensating them using an SLM. 
This not only ensures accurate, interpretable measurements, 
but also significantly reduces the time needed to change the sample and carry out the measurement, in just a few minutes.
}

We first measure the TM of a $24.5$ cm segment of a straight 
$50$-micron core graded-index fiber
(BendBright OM4~\cite{bendbright_OM4}). 
\red{
The choice of fiber studied is guided by the fact that graded-index fibers are the standard for MMFs in telecommunications and are appealing for endoscopic imaging applications due to their relative robustness to bending~\cite{BoonzajerFlaes2018robustness}.
}
Utilizing a fast digital micromirror modulator and an InGaAs camera,
we follow the procedure detailed in Ref.~\cite{matthes2021learning},
which enables us to identify and compensate for aberrations and misalignments.
It also allows us to accurately generate the input masks on the modulator that correspond to rotating the field
with respect to the optical axis in the input facet plane of the fiber.
The principle of the experiment is depicted in Fig.~\ref{fig:setupSimple}
and is further detailed in Appendix~\ref{Appendix:Aberration}.

\begin{figure}[ht]
\includegraphics[width=0.97\columnwidth]{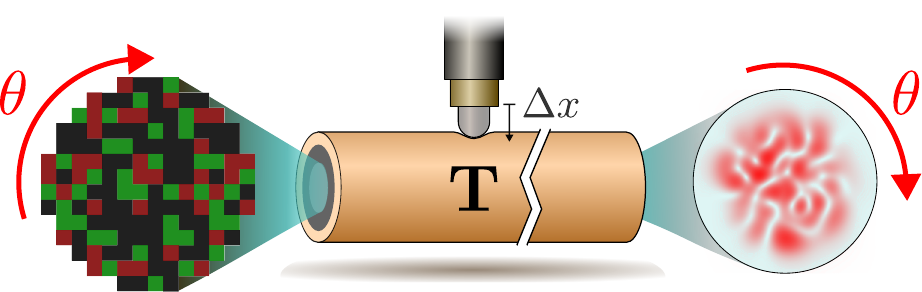}
\caption{
    \textbf{The rotational memory effect in MMF.}
    When the fiber is illuminated by a coherent wavefront $\ket{\psi_{in}}$,
    a seemingly random transmitted field ${\mathbf{T}}\ket{\psi_{in}}$
    is observed at the output.
    In an ideal MMF with cylindrical symmetry, 
    rotating the input wavefront (\textit{i.e.}, sending ${\mathbf{R}(\theta)}\ket{\psi_{in}}$)
    and measuring the transmitted field ${\mathbf{T}\mathbf{R}(\theta)}\ket{\psi_{in}}$,
    is equivalent to rotating the output field resulting from the propagation of $\ket{\psi_{in}}$ and measuring
    ${\mathbf{R}(\theta)\mathbf{T}}\ket{\psi_{in}}$.
    A local perturbation  is then added by moving a tip in contact with the fiber over a distance $\Delta x$ transverse to the fiber axis.
}
\label{fig:setupSimple}
\end{figure}

\subsection{Measurement of the RME}
\label{Sec:MeasurementRME}

\begin{figure*}[t]
\includegraphics[width=0.8\textwidth]{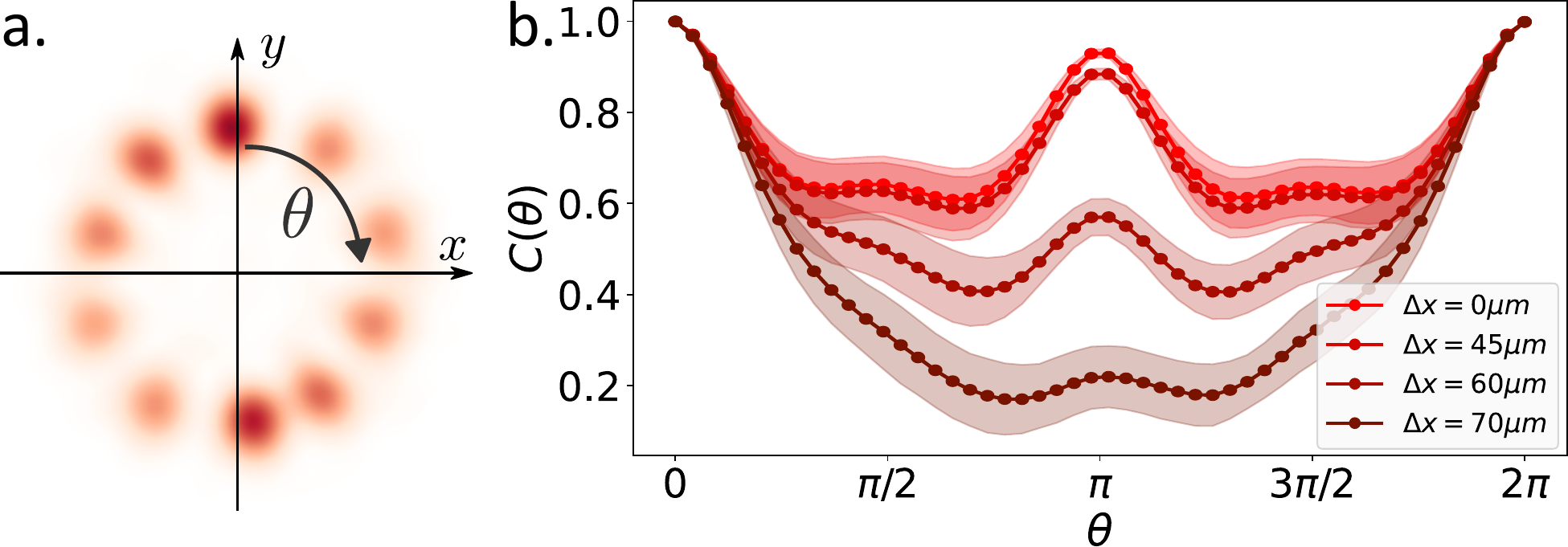}
\caption{
\textbf{Experimental measurement of the RME.}
    (a) Rotation of a focal spot.
        Light is focused at a given output position
        and the input phase mask is rotated along the axis of the fiber
        for 10 values of the rotation angle $\theta$.
         We sum all the resulting output amplitude patterns to reveal
        conservation of the focal spot, albeit with a variation in
         intensity, the latter being
         maximal for angle $\theta=0$ and $\theta=\pi$.
    (b) Experimental measurement of the RME angular correlation~\eqref{eq:corr_function} as a function of the level of perturbation $\Delta x$. 
        The bright red curve shows results for the unperturbed fiber ($\Delta x =0\,\mu$m).
        The red to brown curves correspond to progressively increased disturbances, obtained by applying a local deformation using a translation stage. 
        Results are obtained by averaging over 100 random inputs.
        Shaded areas correspond to the error estimated by the standard deviation of the experimental data.
}
\label{fig:correlation}
\end{figure*}

To illustrate the effect of the RME, 
we first observe its impact on a focusing operation. 
We compute the mask that focuses light at a specific position in the output facet of the fiber 
using the TM~\cite{Popoff2010Measuring,Cizmar2011shaping}. 
It is noteworthy that the knowledge of the TM is not necessary for this step, 
nor for any of our measurements, 
as focusing can be achieved through methods
like sequential optimization~\cite{vellekoop2007focusing,Cizmar2010insitu} 
or phase conjugation~\cite{papadopoulos2012focusing}.
We then rotate the input wavefront.
We show in Fig.~\ref{fig:correlation}(a) the sum of the resulting output amplitude patterns 
for 10 values of the rotation angle. 
We can see that rotating the input masks allows the focusing spot to be rotated along the optical axis of the fiber, with limited degradation of focusing quality.\\

To further characterize the RME, we seek to quantify the similarity 
between a transmitted field 
$\ket{\psi}=\mathbf{T}\ket{\psi_\text{in}}$ 
for a normalized input field $\ket{\psi_\text{in}}$, 
and the output field 
$\ket{\psi_\theta}= \mathbf{T}_\theta\ket{\psi_\text{in}}$, 
where $\mathbf{T}_\theta = \mathbf{R(-\theta)}\mathbf{T}\mathbf{R(\theta)}$. 
The second field corresponds to a rotation of the input and output profiles by an angle of $\theta$ and $-\theta$ respectively. 
We define a correlation function for this purpose as 
\begin{equation}
    C(\theta) = 
        \frac{\vert \ps{\psi}{\psi_\theta} \vert}
        {\sqrt{\ps{\psi}{\psi}}
         \sqrt{\ps{\psi_\theta}{\psi_\theta}}
        }.
    \label{eq:corr_function}
\end{equation}

In practice, we send a set of $100$ random input wavefronts, 
rotate them in the plane of the input facet,
and measure the output field.
We then compute the average correlation function $\left\langle C(\theta)\right\rangle$. 
Figure~\ref{fig:correlation}(b) 
shows the experimentally measured $\left\langle C(\theta)\right\rangle$ 
for the unperturbed fiber (solid red line).
\red{
A first observation is that the correlation deviates from $1$ 
and that the rotated focal spot is degraded 
within the $2\pi$ range.
This indicates the presence of imperfections in the fiber 
that break the cylindrical symmetry of the system.
}
We emphasize that the TM measurement is not used to characterize the memory effect; 
only knowledge of input aberration and misalignment effects is exploited 
to accurately rotate the input field.
Although the results shown here correspond to field correlations, 
we demonstrate in Appendix~\ref{Appendix:IntensityCorr} that very similar results are obtained 
for intensity correlation measurements. The latter can be expressed as $C_I(\theta) \simeq C(\theta)^2$. As a result, the behavior of the angular correlation shown in Fig.~\ref{fig:correlation}(b)   is qualitatively analogous to the result reported in Ref.~\cite{amitonova2015rotational},
where the intensity RME correlation was measured.
Furthermore, we show in Appendix~\ref{Appendix:TMCorr} that when the TM is known, the RME correlation can be accurately computed without the need for additional measurements.

To \red{
qualitatively observe
}
the effect of perturbations on the RME, 
we gradually apply a controlled deformation to the fiber 
along an axis orthogonal to the propagation direction. 
The fiber is maintained on a V-groove and we press locally on the fiber from the top
with a spherical metallic tip using a motorized translation stage.
The correlation $\left\langle C(\theta)\right\rangle$ as a function of the rotation angle $\theta$ is presented in Fig.~\ref{fig:correlation}(b), for different values of the displacement $\Delta x$ of the tip
 (red to brown curves).

We first observe that, even without applying a local perturbation to the fiber, 
the correlation decreases to approximately $60\%$. 
When the fiber is held straight, 
this effect can be attributed to the presence of defects in the fiber. 
This correlation curve exhibits a second maximum, close to $95\%$, at $\theta=\pi$, 
along with small local maxima at $\theta=\pi\pm\pi/2$. 
These features are indicative of the geometrical defects within the fiber.
Upon applying the deformation, 
we observe that the correlation as a function of $\theta$ decreases globally, 
and the local maxima vanishes. 

\red{
\section{Modeling the RME for Defect Characterization}
}
\red{The design and manufacture of few-mode 
and multimode fibers with minimal mode coupling  for mode division multiplexing in telecommunications
pose significant challenges
~\cite{Puttnam2021spaceDivision}. 
Achieving industry-compatible levels of cross-talk 
remains an elusive goal, 
even for fibers with a low mode count (< 10 modes). 
The application of mode-coupling theory to specific types of disorder
enables the prediction of certain adverse effects 
on telecommunications, 
such as mode-dependent losses and power mode coupling~\cite{Marcuse1975mode, Olshansky1975mode}. 
However, the precise characterization of perturbations in a given fiber is difficult to achieve. 
The common approach is to assume coupling solely between pairs of neighboring modes.
Thus, 
with the goal to minimize mode-coupling-induced cross-talk, 
fiber manufacturers 
focus on solely optimizing the difference in 
effective refractive index among the fiber modes~\cite{gruner2012few,Sillard2011fewMode}.
By overlooking the interplay between the symmetries of the modes and the perturbations, 
this method reveals its limitations when attempting to achieve low levels 
of cross-talk~\cite{Maruyama2017relationship}, 
thus restricting the performance of current systems. 
An accurate disorder model with easily estimable parameters 
would be a crucial asset for designing 
multimode fibers with low mode coupling in the telecommunications industry.}

\red{The acute observed sensitivity 
of the RME to geometrical deformations, 
along with its dependence on the fiber's symmetry, 
indicates that one can derive information about the distributed disorder within MMFs. 
In this section, we present}
a theoretical model whose predictions are compared with experimental observations. 
This model proves capable of predicting all RME behaviors in the presence of the disturbances just described.

\subsection{Model of disorder}

In an ideal MMF, 
due to the axisymmetry of the system, 
perfect RME should be expected, \textit{i.e.} the rotation of a given input wavefront should result in a corresponding rotation of the output wavefront. 
This corresponds to $C(\theta) = 1 $ for all $\theta$.
However, real fibers are rarely perfect 
\red{as demonstrated by the result shown in Fig.~\ref{fig:correlation}}, 
resulting in mode coupling 
that is mainly influenced by the 
geometrical defects of the fibers~\cite{ho2013mode}.
Two main contributions can be identified: 
large radius bends, 
attributable to the geometrical conformation of the fiber,
and minor distortions at the core-cladding interface, 
primarily due to fabrication inaccuracies~\cite{Marcuse1973coupled, Mazumder2004analysis, bsaibes2023coupling}.

We propose to model fiber disturbances by 
a deviation of the refractive index profile from a perfect axisymmetric function of the following form:
\begin{equation}
    \delta n(r,\phi, z) =g(z, r) \sum_q \Gamma_q\cos(q\phi + \varphi_q) \, ,
    \label{eq:disorder}
\end{equation} 
where $z$, $r$, and $\phi$ are the cylindrical coordinates 
corresponding respectively to the longitudinal (axis of the fiber), radial, and azimuthal directions. 
The longitudinal variation of $g(z, r)$ are characterized by random fluctuations with 
a correlation length $l_z$, 
which is typically of the order of $100$~\textmu m \red{\cite{bsaibes2023coupling}}, while
radial variations of $g(z, r)$ are discussed in detail in below.
On the other hand, disorder in the azimuthal direction is decomposed into harmonics with orbital momentum $q$ and weight $\Gamma_q$~\cite{Yaxin2022enforcing}.

We attribute the radial fluctuations to variations between neighboring radial layers, 
stemming from inaccuracies in the deposition technique 
or interlayer diffusion of the doping elements. 
Specifically, in the case of the fiber under investigation, 
these inaccuracies are associated with the PCVD process.
This leads us to approximate the fiber of length $L$ by a succession of $N_z = L/l_z $ segments, each of length $l_z$,
in which the perturbation term $\delta n$ is invariant along $z$. 
Specifically, 
for the $p^\text{th}$ segment in the interval $z \in \left[ p\,l_z, (p+1)l_z \right]$, we write
$
\delta n_p(r,\phi) = g_p(r)\sum_q \Gamma_q\cos(q\phi + \varphi_q)
$.
We model $g_p(r)$ as a Gaussian random variable with zero mean, characterized by a standard deviation 
$\sigma_g(r) = d_\text{layer}\vert dn_0(r)/dr\vert$~\cite{Lydtin1986pcvd},
where $n_0(r)$ is the radial profile of the unperturbed fiber, and $d_\text{layer} 
\simeq 10$~nm is the typical length of each layer formed in the PCVD process~\cite{geittner1989manufacturing}.
For a gradient index fiber with a parabolic index profile (see Appendix~\ref{Appendix:HamiltonianTM} and Ref.~\cite{marcuse2013theory}), the standard deviation can be put in the form
\begin{equation}
    \sigma_g(r) \simeq  \frac{r}{a^2}\frac{\text{NA}^2}{ n_\text{max}} d_\text{layer}\mathcal{H}(r/a) \, ,
    \label{eq: sigmag}
\end{equation}
where NA and $a$ are respectively the numerical aperture and radius of the MMF, $n_\text{max}$ is the value of refractive index $n_0(r)$ at the center of the core, and $\mathcal{H}$ is the Heaviside function. 

As detailed in Appendix~\ref{Appendix:HamiltonianTM}, the TM of the $p^\text{th}$ segment of length $l_z$
expressed in the unperturbed fiber mode basis
can be written as

\begin{equation}
    \mathbf{T}_p = e^{-i \left(
                            \mathbf{H}_0 + \mathbf{V}
                        \right) l_z
                     } 
    .
    \label{eq:hamiltonian}
\end{equation}
Here, $\mathbf{H}_0$ is the propagation operator in the absence of perturbation; it is a diagonal matrix containing the propagation constants $\beta_\mu$ of the modes of the unperturbed fiber, indexed by $\mu$. On the other hand,
$\mathbf{V}$ represents the perturbation due to the index fluctuations, $\delta n_p(r,\phi)$,
projected onto the mode basis (see Appendix~\ref{Appendix:HamiltonianTM} for further details). The complete TM is obtained by multiplying the TMs of all the segments.

 \subsection{Theoretical predictions for \texorpdfstring{$\langle C(\theta) \rangle$}{⟨C(θ)⟩}}

In the limit of moderate disorder, we are able to find an analytical expression of the mean correlation function $ \langle C(\theta) \rangle$, which involves the geometrical parameters of the fiber as well as the disorder strength. In Appendix~\ref{Appendix:Theory}, we show that it can be put in the form $ \langle C(\theta) \rangle = \tilde{C}(\theta)/\tilde{C}(0)$,
with
\begin{equation}
   \tilde{C}(\theta) =
            1+ 
            A\sum_{q,\nu,\mu} \Gamma_q^2 
             \cos(q\theta)             B_{\nu\mu}^q\, .
    \label{eq:theo1}
\end{equation}
The prefactor $A=N_z (kl_z)^2/4 N_\text{modes}$ is a coefficient that combines properties of the radial disorder with the number of propagating modes supported by the fiber.
In addition, the coefficient $B_{\nu\mu}^q$ characterizes the energy coupling between  eigenstates 
$\psi_\nu$ and $\psi_\mu$  of the unperturbed propagation operator $\mathbf{H}_0$. 
It is expressed as
\begin{equation}
   B_{\nu\mu}^q = 
        \delta_{m_{\mu \nu},q} \;
        \text{sinc}\left(\frac{\beta_\mu-\beta_\nu}{2}
    l_z
    \right)^2
        I_{\nu\mu} \, ,
    \label{eq:coupling}
\end{equation}
where $m_{\mu \nu} = \vert m_\mu-m_\nu \vert$ is the difference between orbital angular momentum of the eigenstates coupled by the azimuthal disorder, and 
\begin{equation}
I_{\nu\mu} = d_\text{layer}
    \int_0^\infty dr 
        \left|\psi_\nu(r)\right|^2 
        \left|\psi_\mu(r)\right|^2 
        \sigma^2_g(r) r^2 \,
\end{equation}
is the coupling term induced by disorder along the radial direction.

The expression~\eqref{eq:theo1} is a perturbative result, valid when photons scatter on average once over the disordered potential $\mathbf{V}$. As the extent of disorder increases, it becomes necessary to take higher-order perturbations into account. 
This means taking into account multiple interactions between photons and disorder. For all the results presented in this work, the single scattering contribution~\eqref{eq:theo1} is dominant, but we have also calculated the second-order perturbation contribution to obtain quantitative agreement with experimental results and simulations. The second-order contribution takes the following form 
\be
\tilde{C}^{(2)}(\theta)=\tilde{A}\sum_{\substack{q,q'\\ \nu, \kappa, \mu}} \Gamma_q^2\Gamma_{q'}^2 \text{cos}[(q+q')\theta]C_{\nu \kappa \mu}^{qq'},
\label{eq:theo2}
\ee
where $\tilde{A}= N_z (kl_z)^4/16 N_\text{modes}$. Energy coupling is provided by the term
\be
C_{\nu \kappa \mu}^{qq'}= \frac{N_z-1}{2} B_{\nu \kappa}^{q}B_{\kappa \mu}^{q'} +  \delta_{m_{\mu\kappa},q} \delta_{m_{\kappa\nu},q'} Q_{\mu \kappa \nu} I_{\nu \kappa}I_{\kappa \mu},
\ee
where the explicit but lengthy expression of the coefficient $Q_{\mu \kappa \nu}$ in terms of the propagation constants $\beta_\mu$, $\beta_\kappa$, and $\beta_\nu$ is given in Appendix~\ref{Appendix:Theory}. 

\subsection{Validation of the model of disorder and theory}

\begin{figure}[ht]
\includegraphics[width=0.47\textwidth]{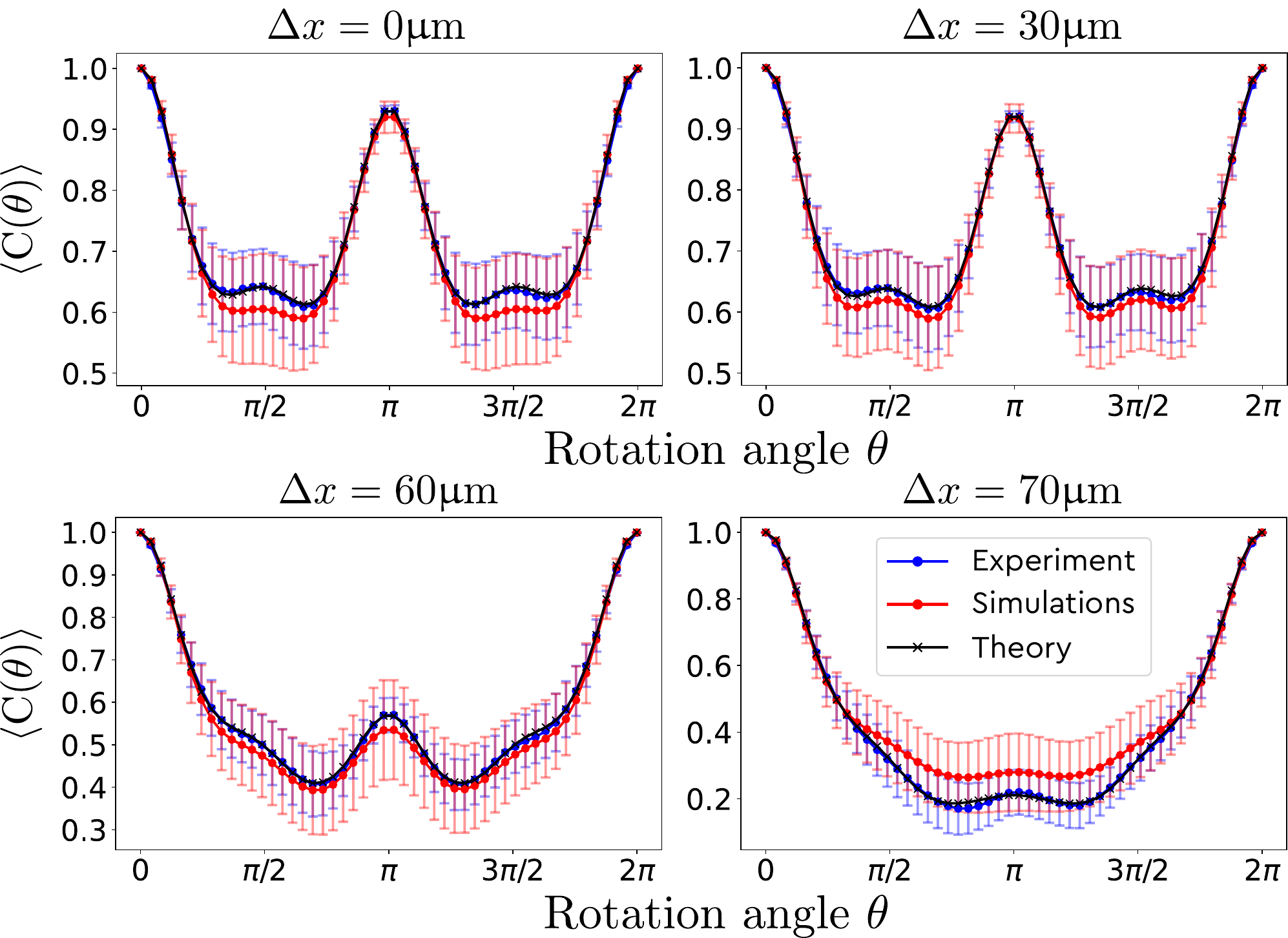}
\caption{
    \textbf{Comparison between experiment, simulations, and theory.}
    Mean angular correlation function of the RME, as defined in Eq.~\eqref{eq:corr_function},
    for various levels of deformation. 
    The fiber used is a typical graded index fiber (Prysmian BendBright OM4~\cite{bendbright_OM4}), 
    with radius $a = 50$~\textmu m, NA$= 0.2$, and $N_\text{modes}=  55$. The correlation length in the model and simulations is set at $l_z = 100$~\textmu m.
    Experimental data (blue lines) are compared 
    with theoretical predictions  based on Eqs.~\eqref{eq:theo1} and~\eqref{eq:theo2} (black lines), 
    and simulation results for wave propagation inside disordered MMFs (red lines). 
    The parameters $\Gamma_q$ of the model are found by fitting to the experimental results, and simulations are obtained with the same parameters. 
    Error bars represent the standard deviation computed over 100 random input wavefronts for the simulations and experiments, as well as 
     20 disorder realizations for the simulations. 
}
\label{fig:theoryVSexpVSsimu}
\end{figure}

To validate the model of disorder as well our theoretical predictions based on Eqs.~\eqref{eq:theo1} and~\eqref{eq:theo2}, we first find the values of the coefficients $\Gamma_q$ that match best the experimental profile of the mean correlation function $\langle C(\theta) \rangle$. 
For the graded-index fibers used in our experiments, we find that it is sufficient to use only four non-zero coefficients corresponding to $q \in [1,2,3,4]$. The experimental results shown in Fig.~\ref{fig:theoryVSexpVSsimu} (blue solid lines) are virtually indistinguishable from the analytical curves (black solid lines).
To assess the physical relevance of these coefficients, 
we then perform simulations using the same values of $\Gamma_q$, without adding any fitting parameter. 
The simulation consists in dividing the fiber into segments of length $l_z$. 
For each segment, we add to the index profile matching the specifications of the fiber
a perturbation of the form given by Eq.~\eqref{eq:disorder}. 
We then estimate its TM using a custom fiber solver~\cite{pyMMF}. 
 Finally, the complete TM of the fiber is obtained by multiplying the TMs of all segments, each segment corresponding to a different realization of the radial disorder (see Appendix~\ref{Appendix:HamiltonianTM} for more details). We calculate the RME correlation as a function of angle and average over 20 realizations of the fiber. The results, presented in Fig.~\ref{fig:theoryVSexpVSsimu}, show good agreement between simulations and theory, with no adjustment parameters required.

\begin{figure}[ht]
\includegraphics[width=0.47\textwidth]{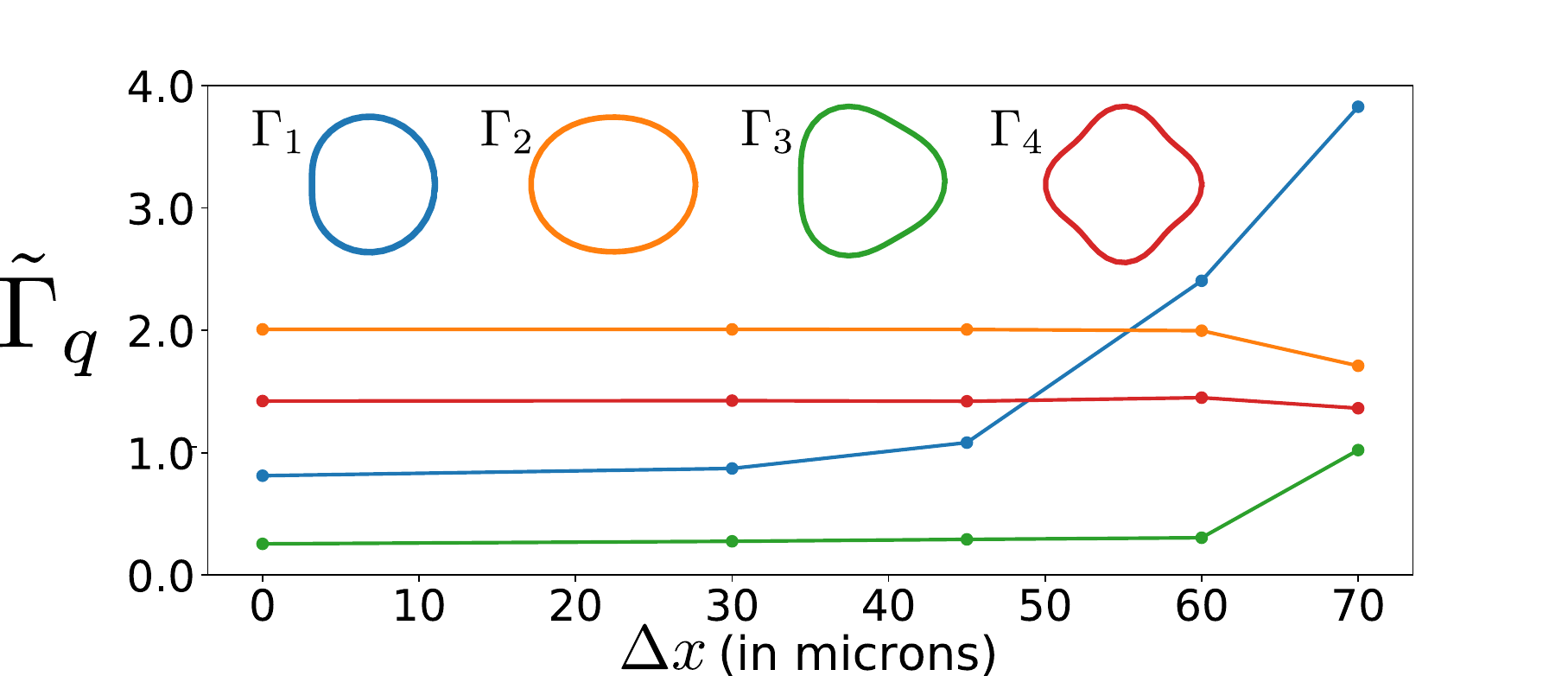}
\caption{
    \textbf{Influence of deformation on the perturbation contributions.}
    Values of the normalized deformation parameters 
    $\tilde{\Gamma}_q = k l_z \sigma_g(r=a) \Gamma_q $.
    The values of $\Gamma_q$ 
    are found by fitting
    the theoretical model [Eqs.~\eqref{eq:theo1} and~\eqref{eq:theo2}] to the experimental data as a function of the deformation.
    In the inset, we show the symmetry corresponding to the perturbation associated with each value of $q$.
}
\label{fig:Gammas}
\end{figure}

\subsection{Discussion and interpretation}

The different values of the angular momenta $q$ of the deformation have different
origins and impacts on the RME. 
Global radial index variations, 
corresponding to $q = 0$, alter the shape of the modes but do not break the axisymmetry of the system. 
\red{
In this case, their effect can simply be equated to a change 
in the length of the fiber up to first order~\cite{BoonzajerFlaes2018robustness}
}
that only impacts the relative phase between the modes.
The TM remains diagonal in the mode basis and commutes with the rotation operator $\mathbf{R}(\theta)$ for any angle $\theta$.
 As a result, the $q=0$ component does not impact the RME. 

In the absence of external perturbation ($\Delta x = 0$), 
thanks to prior compensation for aberrations, 
the deviation of the correlation curve from a perfect RME ($C(\theta) = 1$) 
is due to intrinsic fiber defects caused by the fabrication process,
which give rise to non-zero $\Gamma_q$, for $q>0$. 
In this regime, we find that the correlation function is dominated by the contributions of even values of $q$. 
The contribution $q = 2$ is responsible  for the valleys found at $\theta = \pi \pm \pi/2$, 
and the contribution $q = 4$ for the valleys observed at $\theta = \pi/2 \pm \pi/4$ and $\theta = 3\pi/2 \pm \pi/4\). 
Even contributions have no impact on the value of the correlation at $\theta=\pi$, 
simply because they correspond to $\pi$-symmetric deformations (see inset of Fig.~\ref{fig:Gammas}). 
Consequently, the slight decrease in correlation at $\theta=\pi$ is entirely controlled by the odd deformations.

Although all $\Gamma_q$ terms are of the same order of magnitude, 
we observe that the effect of odd contributions, which couple modes of different parity to the orbital angular momentum, 
is much less pronounced than even contributions, which couple modes of the same parity. 
This is explained by the modal properties of the fiber. 
Indeed, in ideal graded-index fibers,
modes in quasi-degenerate groups, 
i.e. with similar propagation constants $\beta_\mu$,
have the same parity of the angular orbital momentum.
This property is inherited from the modes of the two-dimensional isotropic harmonic oscillator which represents the idealized parabolic graded-index fiber with no boundary~\cite{VANENK1993geometric,gutierrez2019generalized,gutierrez2020majorana,gutierrez2023exactly}. 
Consequently, for pairs of modes for which $m_{\mu\nu}$ is odd, 
the difference of propagation constants $\beta_\mu-\beta_\nu$ is non-negligible, 
leading to weak contributions of $B_{\nu \mu}^q$ appearing in Eq.~\eqref{eq:theo1}. 
This effect has the same origin as 
the observation that disorder preferentially induces coupling between degenerate modes~\cite{matthes2021learning}.

The previous analysis fully explains the robustness of the correlation observed at $\theta =\pi$ for small deformations. 
We note that this correlation revival can equivalently be interpreted in terms of commutation  between the matrix $\mathbf{T}$ and the rotation matrix $\mathbf{R(\theta)}$. 
At small deformations, $\mathbf{T}$ is block-diagonal, with blocks corresponding to quasi-degenerate modes. 
Since, within each block, the different angular momenta $m_\mu$ share the same parity 
(corresponding to constant values of $\vert m_\mu\vert+2p_\mu$, 
where $p_\mu$ is the radial index~\cite{Olshansky1975mode}),
the expression of $\mathbf{R}(\theta)$ restricted to each block necessarily satisfies $\mathbf{R}(\theta=\pi) = \pm \mathbb{1}$, where $\mathbb{1}$ is the identity matrix. As a result $\mathbf{R(\theta=\pi)}$ and $\mathbf{T}$ commute, regardless of the coupling complexity within each group of modes.

When the external mechanical deformation is introduced,
we find that the value of $\Gamma_1$,
corresponding to a flattening of the fiber,
gradually increases (see Fig.~\ref{fig:Gammas}). 
This means that modes of different propagation constant become more and more coupled, 
and that the TM progressively loses its block-diagonal structure in the mode basis.  This explains the disappearance of the dominant revival effect at $\theta =\pi$, as well as the loss of the local maxima at  $\theta =\pi/2$ and $\theta =3\pi/2$. As $\Delta x$ is further increased, we also observe that higher harmonics start to play a role (see $\Gamma_3$ in Fig.~\ref{fig:Gammas}), 
and the width of the correlation function $\langle C(\theta)\rangle$ starts to decrease. 
This indicates that the width of the correlation function at large deformation depends on the disorder strength and is intimately connected to the loss of the block-diagonal structure of the TM.

In Appendix~\ref{Appendix:GRIN}, we present measurements of $\langle C(\theta)\rangle$ obtained for various graded-index fibers, which exhibit advertised properties similar to those of the fiber used in Figs.~\ref{fig:theoryVSexpVSsimu} and~\ref{fig:Gammas}.
 Although we obtain qualitatively similar results, we do find some quantitative reproducible differences, 
expressed in terms of different values for the $\Gamma_q$ weights.      
This demonstrates that RME is a very good indicator for probing the small variations in disturbances that occur during the MMF manufacturing process.\

\red{
The possibility to characterize the distributed perturbation along a fiber using intensity measurements from the input and output facets 
represent important prospects for telecommunication applications. 
Indeed, once the parameters of the perturbations ($l_z$, $\Gamma_q$) are found from a fit of the measured RME correlation, 
one has access to a more accurate model of the fiber.
This model can then be exploited through simulations or mode-coupling theory computations
to predict the cross-talk and losses given the estimated index profile.
This can be used for improving the design and the characterization of new optical fibers 
with desired properties.}

\red{
\section{Improving the RME for imaging prospects}
}

\red{
The potential application of the RME extends to 
the possibility of facilitating blind imaging 
through an unknown MMF, 
in a manner similar to demonstrations in scattering media~\cite{bertolotti2012non-invasive,katz2014non-invasive}. 
For successful image reconstruction, 
it is essential to collect information from the output facet, 
which necessitates maintaining a high RME correlation 
throughout the entire $2\pi$ range. 
As shown in the previous section, 
the memory effect is affected by the presence of disorder 
stemming from the fabrication process. 
Notably, this disorder degrades the RME correlation, 
particularly reducing the range over which 
the correlation remains close to 1.
However, this correlation was 
}
obtained by averaging over random input wavefronts. We now ask whether it is possible to find specific input wavefronts for which the correlation is significantly higher than the mean value for one given angle or for a wide range of angles. 
As with the approaches to tailoring the angular memory effect in scattering media~\cite{yilmaz2021customizing}, we can build operators whose eigenstates optimize the memory effect at a given angle. 
Since losses are low in the fiber, TM is close to unitary and 
the lower part of Eq.~\eqref{eq:corr_function}
is approximately constant.
Then, an interesting operator is the one involved in the upper part of the correlation function,

\begin{equation}
    \mathbf{O}(\theta_t) = \mathbf{T}^\dagger\mathbf{R}(\theta_t)^\dagger\mathbf{T}\mathbf{R}(\theta_t).
    \label{eq:RME_op}
\end{equation}
\begin{figure}[ht]

\includegraphics[width=0.47\textwidth]{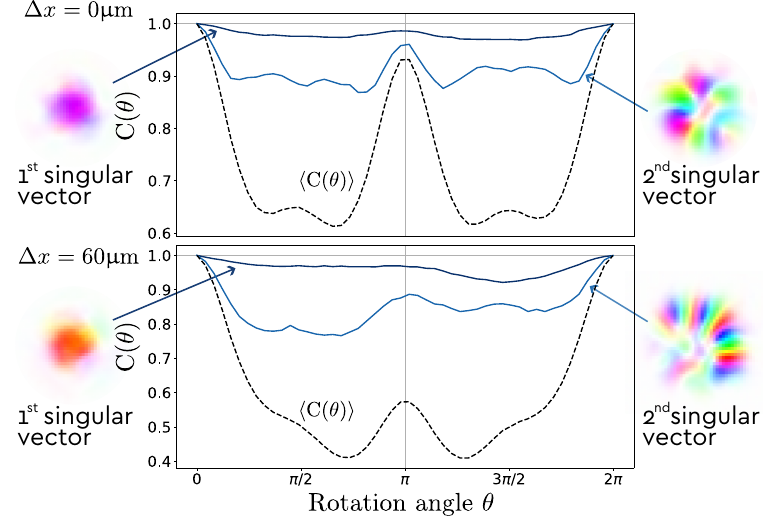}
\caption{
\textbf{Tailoring the rotational memory effect.}
The angular correlation function $C(\theta)$ is constructed using experimentally measured input channels with 
improved RME range, for two values of the deformation ($\Delta x=0\,\mu$m and $\Delta x=60\,\mu$m). The results for the first two  singular vectors of the operator defined in Eq.~\eqref{eq:RME_sum_op}  are compared with the average results for random input profiles (dashed line). 
The insets show the output spatial transverse profiles of the corresponding singular vectors. 
}
\label{fig:tailoring}
\end{figure}

This operator can thus be used to improve the RME for a specific value $\theta_t$ of $\theta$, 
as shown in Appendix~\ref{Appendix:OperatorOneAngle}. 
In order to improve the correlation over the entire $2\pi$-range, 
we can also study the operator built using the sum of operators describing the correlation at different angles,
\begin{equation}\mathbf{O}_\text{sum}=\sum_t\mathbf{T}^\dagger\mathbf{R(\theta_t)}^\dagger\mathbf{T}\mathbf{R(\theta_t)} .
    \label{eq:RME_sum_op}
\end{equation}

To optimize the RME correlation, we construct this operator using the experimentally measured TMs 
with $\theta_t = t\cdot\pi/4$, where $t\in[0, 7]$.
We then compute the singular vectors of this operator corresponding to the singular values 
with the largest modulus. 
We present in Fig.~\ref{fig:tailoring} the resulting correlation $C(\theta)$ for 
the two first eigenvectors 
in the case of no deformation and strong deformations ($\Delta x = 60$~\textmu m) 
and the corresponding output field profiles.
Our results demonstrate that it is possible to find input wavefronts for which the output profiles remain highly correlated across the entire $2\pi$-range.

\red{
Compared previous works that aim to enhance the robustness of an input channel against the level of disorder using a Wigner-Smith operator~\cite{matthes2021learning},
our approach does not depend on differentiating the TM, 
allowing us to study a fixed realization of disorder. 
Furthermore, the projection of this operator onto a given input wavefront 
directly represents the targeted quantity, i.e. the correlation of the RME 
for this specific wavefront.
}

An interesting application of memory effects is the ability to retrieve information from the distal side, where the field for a given input wavefront is \textit{a priori} unknown. 
For imaging applications, 
the range of the memory effect must be wide enough to cover the size of the object to be imaged, 
and the output excitation must have a pronounced peak autocorrelation function~\cite{bertolotti2012non-invasive}. 
\red{Indeed, 
even if the precise shape of the output pattern is unknown, 
approximating its autocorrelation by a Dirac function in space 
allows for accessing the autocorrelation of the hidden object. 
With this information, 
the image of the object can then be estimated 
using a numerical process.
}
This condition is guaranteed in multiple scattering media by the presence of strong disorder that randomizes the field 
for any given input wavefront. However, this is not the case in MMFs, where the disorder does not affect all modes in the same way~\cite{Cizmar2011shaping}.
A trivial solution for maximizing the RME range would be to use the fundamental mode, 
which is less affected by external perturbations~\cite{matthes2021learning}. But, due to its rotational symmetry, the autocorrelation of this mode with respect to angular rotation is close to one. 
So, even though the field profile remains correlated at the fiber's output when the input profile is rotated, 
this mode cannot be used to provide information about the distal end of the fiber. 
As shown in Fig.~\ref{fig:tailoring}, the first singular vector of the operator~\eqref{eq:RME_sum_op} is very close to the fundamental mode for any $\Delta x$ and is therefore not useful for imaging. However, the second combines the properties of a large-range RME and an output pattern with a peaked autocorrelation function (see Appendix~\ref{Appendix:AutoCorrelation} for details). It is therefore a good candidate for recovering information about the fiber distal end.

\section{Conclusion}

In this article, we first present an approach based on the TM measurement that enables us 
to accurately measure and study RME in MMFs.
Importantly, this method allows us to mitigate the effects of aberrations and misalignments 
that can significantly disrupt RME analysis.

We then propose a model of disorder and provide  a theoretical calculation  of the RME correlation function, which is shown to be in good agreement with both experimental data 
and \red{realistic wave propagation} simulations.
In particular, our analysis makes it possible to estimate geometric perturbations in the fiber, 
whether due to fabrication imperfections or mechanical deformations. 
\red{From a fundamental perspective,} 
this approach can serve as a powerful tool for the study of MMF defects 
resulting from the breaking of fiber symmetry. 
\red{
Moreover, the unknown parameters of the disorder along the azimuthal and longitudinal directions 
can be determined through simple input-output measurements.}

\red{
This opens up promising avenues for  
designing and characterizing fibers for telecommunications applications, 
aiming to reduce modal cross-talk that currently restricts their practical utility.
}

Finally, 
\red{we tackle the issue of the robustness of the RME with respect to the rotation angle for a given disorder.
This is an important issue for the prospect of 
harnessing the RME for imaging applications.
}
We demonstrate the possibility of generating channels that exhibit
a drastic improvement in the RME. 
In particular, we can create channels that 
are more robust to deformations compared to random inputs or standard fiber modes, 
and that also exhibit a random profile with high spatial frequencies.

\section*{Acknowledgments}

\noindent 
R.G.C, A.G, J.R. and S.M.P acknowledge the French \textit{Agence Nationale pour la Recherche} grant No. ANR-23-CE42-0010-01 MUFFIN 
and the  Labex WIFI grant No. ANR-10-LABX-24, ANR-10-IDEX-0001-02 PSL*. 
R.G.C, A.G., E.R.A., L.B., Y.Q., M.B., P.S., M.B., J.R., and S.M.P acknowledge the French \textit{Agence Nationale pour la Recherche} grant No. ANR-20-CE24-0016 MUPHTA.

\section*{Data and code availability}

\noindent Raw and processed data, sources to regenerate the all the figures,
and sample codes for the treatment pre- and post-processing are  available in the dedicated repository~\cite{repo}.

\appendix
%%%%%%%%%%%%%%%%%%%%%%%%%%%%%%%%%%%%%%%%%%%%%%%%%%%%%%%%%%%%%%%%%%%%
%%%%%%%%%%%%%%%%%%%%%%%%%%% ANNEXE A %%%%%%%%%%%%%%%%%%%%%%%%%%%%%%%
%%%%%%%%%%%%%%%%%%%%%%%%%%%%%%%%%%%%%%%%%%%%%%%%%%%%%%%%%%%%%%%%%%%%
\section{TM and RME measurements}

\subsection{Aberration compensation and TM measurement}
\label{Appendix:Aberration}

To decouple the effects of the RME and measurement inaccuracies, we use the approach we developed 
to learn and compensate for aberrations and misalignments in Ref.~\cite{matthes2021learning}. 
The idea is to first measure the TM on a pixel basis and then project it onto the mode basis.
Without aberrations, this projection into the mode basis should conserve energy, since all the energy must be conveyed by those modes. 
Using a model-based algorithm, constructed with the deep-learning framework PyTorch~\cite{NIPS2019_9015}, 
we identify the aberrations and misalignments of the system that minimize the loss when projecting onto the mode basis.
First, this process enables us to accurately recover the TM in the mode basis, 
which contains all the information about light propagation in the MMF. 
Secondly, it facilitates the identification of the aberrations that need to be corrected 
in order to obtain a desired pattern in the input facet plane of the fiber. 
The correction can then be implemented onto the SLM.

%%%%%%%%%%%%%%%%%%%%%%%%%%%%%%%%%%%%%%%%%%%%%%%%%%%%%%%%%%%%%%%%%%%%
%%%%%%%%%%%%%%%%%%%%%%%%%%% ANNEXE B %%%%%%%%%%%%%%%%%%%%%%%%%%%%%%%
%%%%%%%%%%%%%%%%%%%%%%%%%%%%%%%%%%%%%%%%%%%%%%%%%%%%%%%%%%%%%%%%%%%%
% \section*{Appendix B: Correlation estimation of the RME in different configurations}

\subsection{Intensity vs field correlation}
\label{Appendix:IntensityCorr}
In the main text, we studied the RME using the field correlation function~\eqref{eq:corr_function}. Another way to characterize the RME amounts to estimating the
correlation between the output intensity patterns associated to the fields $\psi(\vec{r})$ and $\psi_\theta(\vec{r})$ involved 
in Eq.~\eqref{eq:corr_function}~\cite{amitonova2015rotational}.
It corresponds the output intensity pattern $I(\vec{r})=\vert \psi(\vec{r}) \vert^2$ for a given input, 
and the intensity $I_\theta(\vec{r})= \vert \psi_\theta(\vec{r}) \vert^2$ obtained when rotating the input and output profiles by an angle $\theta$. The intensity correlation reads

\begin{equation}
    C_I(\theta) = 
    \frac{
        \int d\vec{r} I(\vec{r}) I_\theta(\vec{r})
    }{
        \sqrt{
            \int d\vec{r} I(\vec{r})^2
            \int d\vec{r} I_\theta(\vec{r})^2
        }   
    }.
    \label{eq:corr_function_I}
\end{equation}

\begin{figure}[ht]
\includegraphics[width=0.47\textwidth]{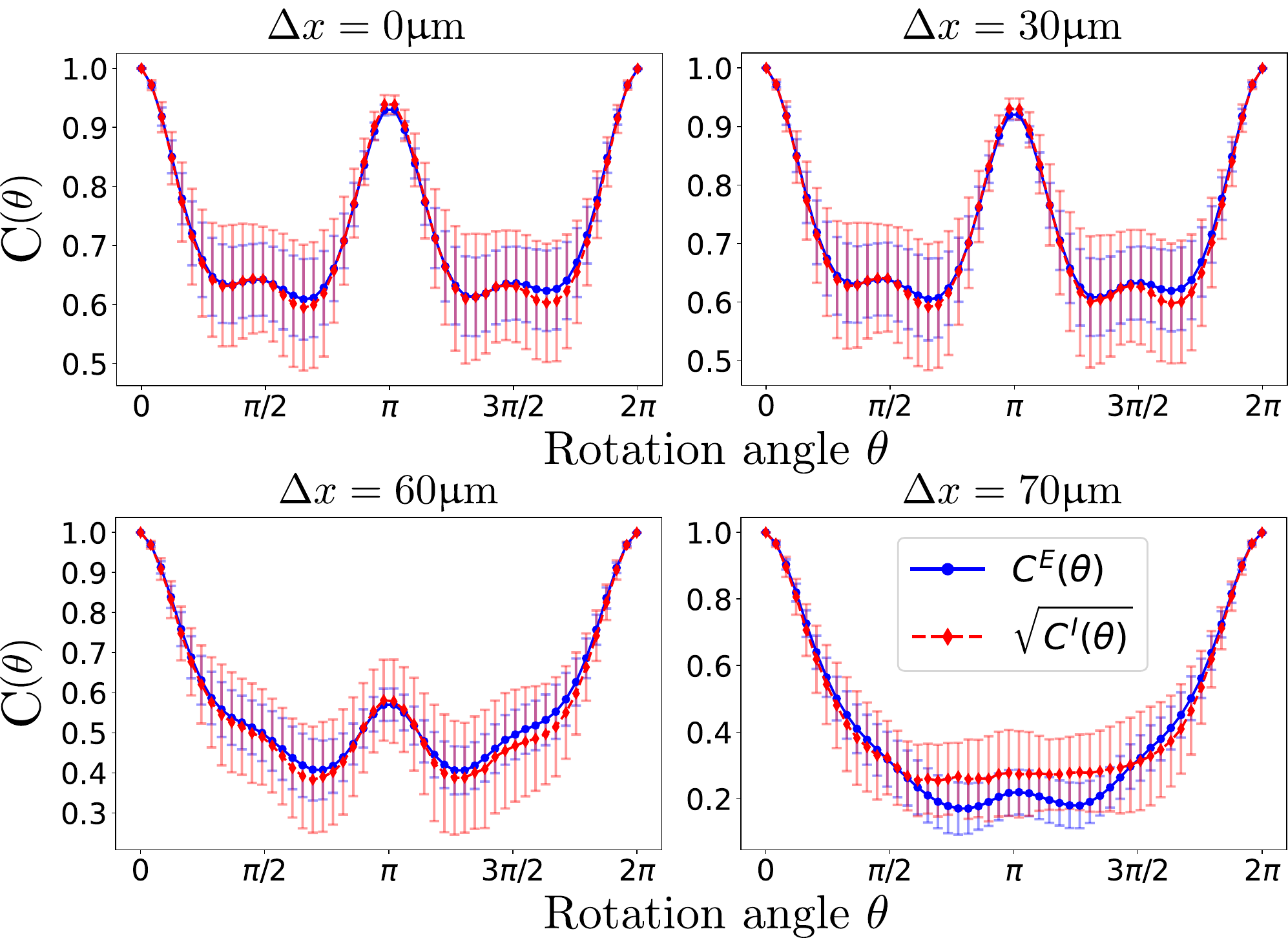}
\caption{
    \textbf{Comparison between the field and the intensity correlations as a function of the rotation angle $\theta$.}
    For different values of the deformation $\Delta x$, 
    we show the field correlation as defined in Eq.~\eqref{eq:corr_function} (blue curve), 
    as well as the square root of the intensity correlation defined in Eq.~\eqref{eq:corr_function_I} (red curve).
}
\label{fig:CEvsCI}
\end{figure}

Such correlation function was originally used in Ref.~\cite{amitonova2015rotational}. To estimate it, we use the same procedure and apparatus as the ones discussed in Sec.~\ref{Sec:MeasurementRME}, where intensities can be evaluated from the field measurements.
We then compare the mean field correlation~\eqref{eq:corr_function}
with the square root of the mean intensity correlation~\eqref{eq:corr_function_I}. 
As illustrated in Fig.~\ref{fig:CEvsCI}, the mean values of the two correlation functions are in excellent agreement. This  demonstrates that the observables $\langle C(\theta) \rangle$ and $\langle C_I(\theta) \rangle$ are equivalent.

\subsection{Estimation of the correlation using the measured TM}
\label{Appendix:TMCorr}

\begin{figure}[ht]
\includegraphics[width=0.47\textwidth]{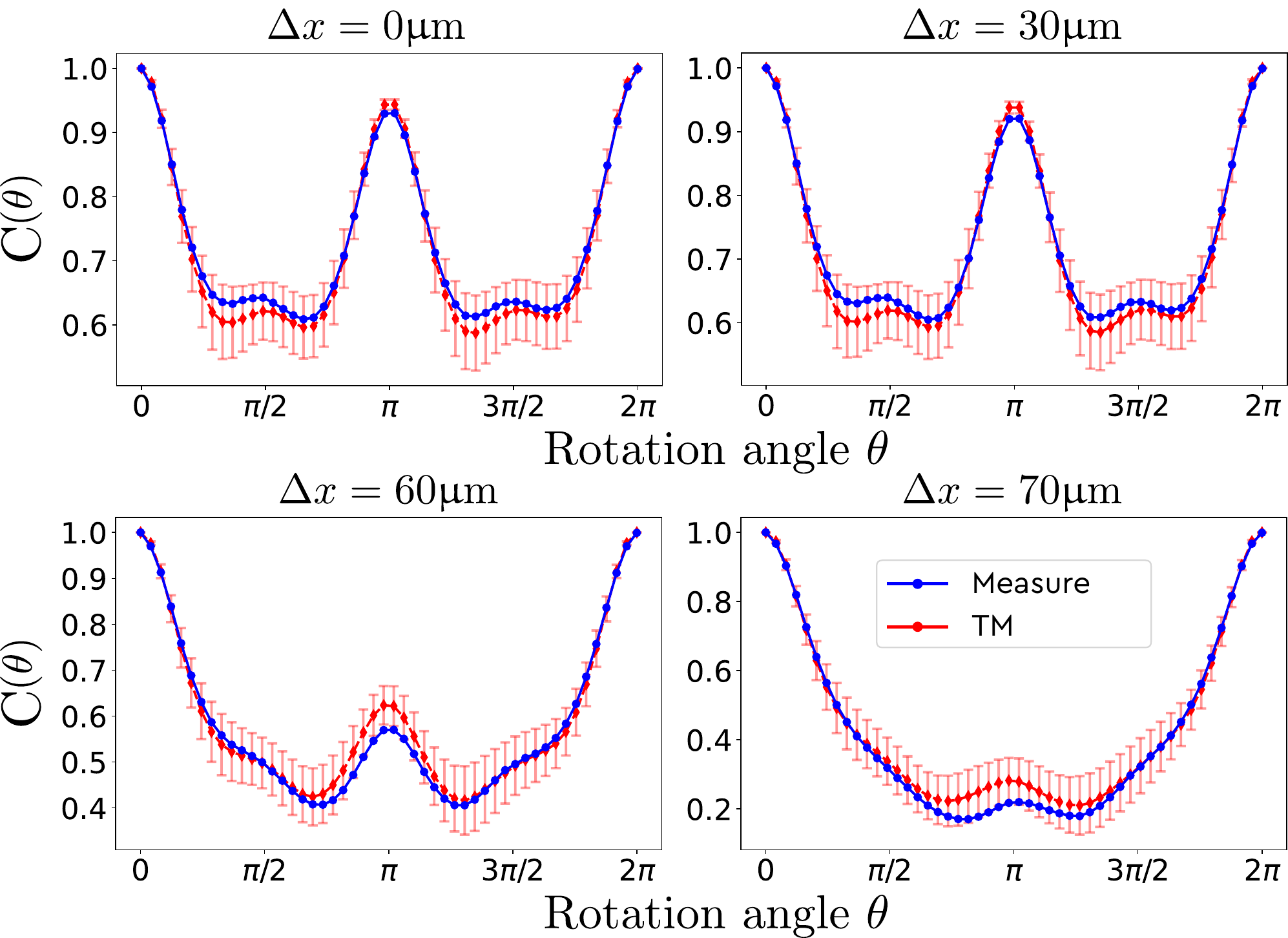}
\caption{
    \textbf{Comparison between the correlation $C(\theta)$ based on the measurement of the output fields, and the one estimated using the TM.}
    For different values of the deformation $\Delta x$, 
    we show the field correlation as defined in Eq.~\eqref{eq:corr_function} (blue curve), 
    and the one obtained using the TM (red curve).
}
\label{fig:TMvsCE}
\end{figure}

In the present study,
we use the compensation of aberrations, facilitated by the transmission matrix (TM) measurement,
but we do not directly employ the knowledge of the TM itself.
However, the TM provides access to the output field $\ket{\psi_{\text{out}}}$
for any given input field $\ket{\psi_{\text{in}}}$.
We can thus use the TM to estimate the output of a rotated wavefront and compute the correlation function defined in Eq.~\eqref{eq:corr_function}.
For each deformation, we compute the mean correlation for 100 random input wavefronts.
We show in Fig.~\ref{fig:TMvsCE} a good agreement between the estimation based on the TM and the one based on the explicit measurement of the output field. 
This demonstrates that the measurement of the TM can drastically reduce
the time needed for characterizing the RME, 
as it does not require any additional measurement. 
In comparison, the explicit measurements presented in the main text necessitate, 
after the initial calibration, 
to average over 100 random input wavefronts for 50 different angles.

%%%%%%%%%%%%%%%%%%%%%%%%%%%%%%%%%%%%%%%%%%%%%%%%%%%%%%%%%%%%%%%%%%%%
%%%%%%%%%%%%%%%%%%%%%%%%%%% ANNEXE B %%%%%%%%%%%%%%%%%%%%%%%%%%%%%%%
%%%%%%%%%%%%%%%%%%%%%%%%%%%%%%%%%%%%%%%%%%%%%%%%%%%%%%%%%%%%%%%%%%%%
\section{Effective Hamiltonian and transmission matrix}
\label{Appendix:HamiltonianTM}
In the situation where the coupling between different polarization channels can be neglected, 
and in the  weakly guiding approximation~\cite{marcuse2013theory} 
(\textit{i.e.} for variations of the index of refraction small compared to the average index value), 
each polarization of the transverse part of the field at frequency $\omega$ satisfies the scalar wave equation
\be
\left[ \nabla_\perp^2 + \partial_z^2 +k^2n(\vec{r}, z)^2 \right]\psi(\vec{r}, z)=0,
\ee
where $k=\omega/c$ and $\vec{r}=(r,\phi)$ labels the position in the transverse plane. The refractive index is further decomposed into an unperturbed axisymmetric component $n_0$ and a perturbation $\delta n \ll n_0$,
\be
n(\vec{r}, z)=n_0(r)+\delta n(\vec{r}, z).
\ee

To identify the effective Hamiltonian that controls the dynamics in the presence of disorder, it is convenient to rewrite the wave equation in the operator form
\be
\partial^2_z\ket{\psi(z)}=-\hat{H}(z)^2\ket{\psi(z)},
\label{EqDynH}
\ee
where 
\begin{align}
\hat{H}(z)&=\left[ \hat{\nabla}_\perp^2 +k^2\hat{n}(\vec{r}, z)^2 \right]^{1/2}
\nonumber
\\
&\simeq \left[\hat{H}_0^2 +2 k^2 \hat{n}_0(r)\delta \hat{n}(\vec{r}, z)  \right]^{1/2}
\nonumber
\\
&\simeq \hat{H}_0+ k^2 \hat{H}_0^{-1}\hat{n}_0(r)\delta \hat{n}(\vec{r}, z).
\end{align}
The Hamitonian of the unperturbed problem reads $\hat{H}_0=\left[ \hat{\nabla}_\perp^2 +k^2\hat{n}_0(r)^2 \right]^{1/2}$. Since in a realistic MMF, the relative variations of $n_0(r)$ in the radial direction are small, the eigenvalues $\beta_\mu$ of $\hat{H}_0$ are close to $kn_0$, where $n_0$ is the typical refractive index of the core. Therefore, a good approximation of $\hat{H}$ is 
\be
\hat{H}(z)\simeq \hat{H}_0 + k\delta \hat{n}(\vec{r}, z).
\label{EqHamiltonianSector}
\ee

In the present work, back reflections can be neglected, and Eq.~\eqref{EqDynH} is equivalent to 
\be
\partial_z\ket{\psi(z)}=-i\hat{H}(z)\ket{\psi(z)}.
\ee
This shows that the the field transmitted through the fiber of length $L$ can be expressed in terms of a unitary transmission matrix $\mathbf{T}$ as $\ket{\psi(L)}=\mathbf{T}\ket{\psi(0)}$. The matrix $\mathbf{T}$ reads
\be
\mathbf{T}= \mathcal{T}e^{-i\int_0^L dz'\hat{H}(z') },
\ee
where $\mathcal{T}$ is the time-ordering operator ($z$ plays the role of time here). In the following, we model the disorder along the propagation direction $z$ as a succession of $N_z=L/l_z$ independent segments of length $l_z$, where the refractive index depends only on the transverse coordinate $\vec{r}$. In that case, the transmission matrix takes the form
\be
\mathbf{T}=\prod_{p=1}^{N_z} \mathbf{T}^{(p)},
\label{EqTProd}
\ee
 with 
 \be
  \mathbf{T}^{(p)}=e^{-i[\hat{H}_0 + k\delta \hat{n}_p(\vec{r})]l_z}.
  \label{EqTSegment}
 \ee
The index fluctuations of each sector $p$ is expressed as the product of a random function along the radial direction and a random function decomposed on the azimuthal harmonics,
 \be
 \delta n_p(r, \phi)= g_p(r)\sum_q\Gamma_q \text{cos}(q\phi + \varphi_q).
 \label{EqRefactiveIndex}
 \ee
 Here, $g_p(r)$ is a Gaussian random variable with zero mean and variance $\langle g_p(r)g_p(r')\rangle=\sigma_g(r)^2d_{\text{layer}}\delta(r-r')$, where $d_{\text{layer}}$ is the thickness of each layer obtained in the chemical vapor deposition process. In addition, the phases $ \varphi_q$ are random independent variables with uniform distribution, added to mitigate the effect of the orientation of the perturbation. 

In this work, we focus on the properties of graded index fibers, where the refractive index $n_0(r)$ takes the form
\be
n_0(r)^2= n_\text{max}^2\left( 1- 2\Delta \frac{r^2}{a^2}\right)
\ee
 in the core of the fiber of radius $a$. Here $\Delta = \left(n_\text{max} - n_\text{cl}\right)/n_\text{max}$, where
$n_\text{cl}$ is the refractive index in the cladding, \textit {i.e.} for $r > a$. In the weakly guiding approximation ($\text{NA} \ll 1$), 
$\Delta \simeq \text{NA}^2/2 n_\text{max}^2$ and the refractive index profile in the core
is well approximated by a parabolic function, $n_0(r) \simeq n_\text{max}(1-\Delta r^2/a^2)$. This yields the explicit expression~\eqref{eq: sigmag} for the amplitude of the radial disorder $\sigma_g(r)$.

 The expressions~\eqref{EqTProd}, ~\eqref{EqTSegment}, and~\eqref{EqRefactiveIndex} are used both in the theoretical treatment developed in Appendix~\ref{Appendix:IntensityCorr} and in the numerical simulations. For simulation purposes, the modes profiles $\psi_\mu$ and propagation constants $\beta_\mu$ of the unperturbed fiber (which are the eigenstates and eigenvalues of $\hat{H}_0$) are computed using the pyMMF package~\cite{matthes2021learning,pyMMF}.  The Hamiltonian~\eqref{EqHamiltonianSector} and transmission matrix~\eqref{EqTSegment} of each sector $p$ is then computed in the basis $\{ \psi_\mu\}$. Finally, the total TM is found by multiplying the TMs of all the segments, as in Eq.~\eqref{EqTProd}.
 Details of the simulations, performed in Python, are available in the dedicated repository~\cite{repo}.

%%%%%%%%%%%%%%%%%%%%%%%%%%%%%%%%%%%%%%%%%%%%%%%%%%%%%%%%%%%%%%%%%%%%
%%%%%%%%%%%%%%%%%%%%%%%%%%% ANNEXE C %%%%%%%%%%%%%%%%%%%%%%%%%%%%%%%
%%%%%%%%%%%%%%%%%%%%%%%%%%%%%%%%%%%%%%%%%%%%%%%%%%%%%%%%%%%%%%%%%%%%
\section{ Analytical predictions for the RME}
\label{Appendix:Theory}

In this appendix, we evaluate the mean correlator $\langle C(\theta)\rangle= \tilde{C}(\theta)/\tilde{C}(0)$, 
where
\be
\tilde{C}(\theta)= \ov{ \ps{\psi}{\psi_\theta} } = 
 \ov{ \bra{\psi_{\text{in}}} \mathbf{T}^\dagger \mathbf{T}_\theta \ket{\psi_{\text{in}}} },
\label{EqCTilde}
\ee
and $\ov{\dots}=\langle \dots \rangle$ stands for the average over different configurations of the disorder.
We first decompose the input field in the mode basis $\{\psi_\mu\}$ of the unperturbed Hamiltonian $\hat{H}_0$,
\be
\ket{\psi_{\text{in}}}=\sum_{\mu=1}^{N}c_\mu \ket{\psi_\mu},
\ee
where $\sum_{\mu=1}^N \vert c_\mu \vert^2=1$.  In the following, we write the unperturbed eigenmodes in the form
\be
\psi_\mu(r,\phi)=\frac{1}{\sqrt{2\pi}}\varphi_\mu(r)e^{im_\mu \phi},
\ee
so that the normalization condition $\ps{\psi_\mu}{\psi_\mu}=1$ reads
\be
\int_0^\infty dr r \vert \varphi_\mu(r) \vert^2=1.
\ee
In addition, we consider random input wavefronts, uniformly distributed over the $N_\text{modes}$ modes of the MMF. Using $\langle c_\mu c_{\mu'} \rangle= \delta_{\mu, \mu'}/N_\text{modes}$, we express the  correlator~\eqref{EqCTilde} as
\be
\tilde{C}(\theta)=\frac{1}{N_\text{modes}}\sum_{\nu, \mu } e^{-i(m_\nu -m_\mu)\theta} \left< \vert T_{\nu \mu} \vert^2 \right>. 
\label{EqCorrelatorVsT}
\ee

We then use the decomposition~\eqref{EqTProd}, where TMs $\mathbf{T}^{(p)}$ are independent of each other, and satisfy $\langle \mathbf{T}^{(p)} \rangle=0$. This gives
\be
\langle \vert T_{\nu \mu} \vert^2 \rangle = \left( \prod_{p=1}^{N_z} \langle   \mathbf{T}^{(p)} \otimes \mathbf{T}^{(p) \dagger} \rangle \right)_{\nu \mu}.
\label{EqTCorrelator}
\ee
In the case of weak disorder ($N_z (kl_z)^2 \left<\delta n^2 \right> \lesssim 1 $), we can evaluate the previous correlator using a perturbative expansion of each matrix $\mathbf{T}^{(p)}$. To obtain an explicit form of the latter, it is more convenient to work with the interaction representation $\mathbf{T}_I(z)=e^{i\hat{H}_0z}\mathbf{T}(z)$ than directly using the expansion of Eq.~\eqref{EqTSegment}. As the  matrix $\mathbf{T}_I(z)$ obeys the equation $\partial_z \mathbf{T}_I(z)=-i \hat{V}_I(z)\mathbf{T}_I(z)$, where $\hat{V}_I(z)=e^{i\hat{H}_0z}\hat{V}e^{-i\hat{H}_0z}$ and $\hat{V}(z)=k\delta \hat{n}(z)$, it can be expanded, up to the second order in $\hat{V}_I$, in the form
\begin{align}
\mathbf{T}_I(z)&= \mathbb{1} -i\int_0^z dz' \hat{V}_I(z')\mathbf{T}_I(z')
\nonumber
\\
&\simeq \mathbb{1} -i\int_0^z \!\! dz' \hat{V}_I(z') - \int_0^z \!\! dz' \!\! \int_0^{z'} \!\! dz'' \hat{V}_I(z') \hat{V}_I(z'').
\label{EqExpansionTI}
\end{align}
Physically, this expansion corresponds to a situation where photons interact at most twice with the disordered potential located in a section of the fiber of length $z$. 
Inside each sector of length $l_z$, the potential $\hat{V}(z)$ is invariant along $z$, so that  integrals in Eq.~\eqref{EqExpansionTI} can be evaluated explicitly. This allows us to find the expression of  $\mathbf{T}^{(p)}= e^{-i\hat{H}_0l_z}\mathbf{T}_I(l_z)$, up to second order in $\hat{V}=k\delta \hat{n}_p$,
\be
T^{(p)}_{\nu\mu}\simeq e^{-i\beta_\mu l_z} \left( \delta_{\nu \mu}+ T^{(p,1)}_{\nu\mu} + T^{(p,2)}_{\nu\mu} \right),
\label{EqTExpansion}
\ee
where
\begin{align}
T^{(p,1)}_{\nu\mu}&= -il_ze^{i\beta_{\nu \mu}l_z/2}\text{sinc}\left(\beta_{\nu \mu}l_z/2\right)V_{\nu \mu},
\\
T^{(p,2)}_{\nu\mu}&=-il_z \sum_{\kappa} \frac{e^{i\beta_{\nu \kappa}}}{\beta_{\nu \mu}}
\left[
e^{i \beta_{\kappa \mu}l_z/2}\text{sinc}\left(\beta_{\kappa \mu}l_z/2\right)
\right.
\nonumber
\\
&\;\;\;\; + \left.  e^{-i \beta_{\nu\kappa}l_z/2}\text{sinc}\left(\beta_{\nu\kappa}l_z/2\right)\right] V_{\nu \kappa}V_{\kappa \mu}
,
\label{EqTOrder2}
\end{align}
with $\beta_{\nu \mu}=\beta_\nu-\beta_\mu$.  Inserting the expansion~\eqref{EqTExpansion} into Eq.~\eqref{EqTCorrelator} and keeping terms up to second order in $V$, we obtain
\begin{align}
\langle \vert T_{\nu \mu} \vert^2 \rangle  \simeq  & \;  \delta_{\nu \mu} + N_z \langle \vert T^{(p, 1)}_{\nu \mu} \vert^2 \rangle  + N_z \langle \vert T^{(p, 2)}_{\nu \mu} \vert^2 \rangle
\nonumber
\\
&+ \frac{N_z(N_z-1)}{2} \sum_\kappa   \langle \vert T^{(p, 1)}_{\nu \kappa} \vert^2 \rangle  \langle \vert T^{(p, 1)}_{\kappa \mu} \vert^2 \rangle.
\label{EqTSolSecondOrder}
\end{align}
First-order contributions are of the form
\be
 \langle \vert T^{(p, 1)}_{\nu \mu} \vert^2 \rangle= l_z^2 \,\text{sinc}\left(\beta_{\nu \mu}l_z/2\right)^2  \langle \vert V_{\nu \mu} \vert^2 \rangle,
\ee
where $V_{\nu \mu}=k \bra{\psi_\nu} \delta\hat{n}_p \ket{\psi_\mu}$. For the model of disorder given by Eq.~\eqref{EqRefactiveIndex}, we find
\be
\langle \vert V_{\nu \mu} \vert^2 \rangle = \frac{k^2}{4} I_{\nu \mu} \sum_q \Gamma_q^2 \delta_{q, \vert m_\nu-m_\nu \vert},
\ee
where 
\be
I_{\nu \mu} = d_{\text{layer}}\int dr  \vert \psi_\nu(r)\vert^2 \vert \psi_\mu(r)\vert^2 \sigma_g(r)^2  r^2.
\ee
Second order contributions $ \langle \vert T^{(p, 2)}_{\nu \mu} \vert^2 \rangle$ involve averages of the form $\mathcal{C}_{\nu \kappa \mu}^{\nu \kappa' \mu} = \langle V_{\nu \kappa} V_{\kappa \mu} V_{\nu \kappa'}^* V_{\kappa' \mu}^* \rangle $, which we can contract as
\begin{align}
\mathcal{C}_{\nu \kappa \mu}^{\nu \kappa' \mu}
&= \langle V_{\nu \kappa} V_{\nu \kappa'}^*\rangle
\langle V_{\kappa \mu} V_{\kappa' \mu}^* \rangle 
+
\langle V_{\nu \kappa} V_{\kappa' \mu}^*  \rangle
\langle V_{\kappa \mu} V_{\nu \kappa'}^*  \rangle
\nonumber
\\
&\simeq \langle \vert V_{\nu \kappa} \vert^2 \rangle \langle \vert V_{\kappa \mu} \vert^2 \rangle \delta_{\kappa \kappa'} + \langle \vert V_{\nu \nu} \vert^2 \rangle^2 \delta_{\kappa \kappa'}\delta_{\nu \mu} \delta_{\nu \kappa}.
\label{EqContractionV4th}
\end{align}
Combining the expression~\eqref{EqTOrder2} with the previous result, we obtain
\be
 \langle \vert T^{(p, 2)}_{\nu \mu} \vert^2 \rangle \simeq l_z^4  \sum_\kappa Q_{\nu \kappa \mu}
 \langle \vert V_{\nu \kappa} \vert^2 \rangle
 \langle \vert V_{\kappa \mu} \vert^2 \rangle,
\ee
where $Q_{\nu \kappa \mu}$ is a coupling weight between different energy subspaces,
\begin{align}
&Q_{\nu \kappa \mu}=\frac{1}{\beta_{\nu \mu}^2l_z^2} \left[
\text{sinc}\left(\beta_{\nu\kappa}l_z/2\right)^2 + \text{sinc}\left(\beta_{\kappa \mu}l_z/2\right)^2
\right.
\nonumber
\\
&\left.-2\text{sinc}\left(\beta_{\nu\kappa}l_z/2\right) \text{sinc}\left(\beta_{\kappa \mu}l_z/2\right)
\text{cos}\left( \beta_{\nu\mu}l_z/2 \right)
\right] + \frac{1}{4}\delta_{\nu \kappa}\delta_{\kappa \mu}.
\end{align}

Finally, we insert the result~\eqref{EqTSolSecondOrder} into the expression~\eqref{EqCorrelatorVsT} of the correlator, to get an expansion of the form
\be
\tilde{C}(\theta)= 1+ \tilde{C}^{(1)}(\theta) + \tilde{C}^{(2)}(\theta).
\label{EqSolExpansion}
\ee
The first order in $V^2$ reads
\be
 \tilde{C}^{(1)}(\theta)= A_1 \!\!\!\!\!\! \sum_{\substack{q,\nu, \mu\\ m_\nu -m_\mu = \pm q}} 
  \!\!\!\!\!
  \Gamma_q^2 \text{cos}(q\theta)  I_{\nu \mu} \text{sinc}\left(\frac{\beta_{\nu \mu}l_z}{2}\right)^2,
\label{EqFirstOrder}  
\ee
where $A_1=N_z (kl_z)^2/4N_\text{modes}$. On the other hand, the second order in $V^2$ reads
\be
 \tilde{C}^{(2)}(\theta)= A_2  \!\!\!\!\!\!\! \sum_{\substack{q, q',\nu, \kappa, \mu \\ m_\nu -m_\kappa=q \\ m_\kappa -m_\mu =\pm q' }} \!\!\!\!\!\! \Gamma_q^2\Gamma_{q'}^2 \text{cos}[(q\pm q')\theta]I_{\nu \kappa} I_{\kappa \mu} \tilde{Q}_{\nu \kappa \mu},
 \label{EqSecondOrder} 
\ee
where $A_2=N_z (kl_z)^4/8N_\text{modes}$, and 
\be
\tilde{Q}_{\nu \kappa \mu}=Q_{\nu \kappa \mu} + \frac{N_z-1}{2}\text{sinc}\left(\frac{\beta_{\nu\kappa}l_z}{2}\right)^2 \text{sinc}\left(\frac{\beta_{\kappa \mu}l_z}{2}\right)^2.
\ee
Equations~\eqref{EqSolExpansion}, ~\eqref{EqFirstOrder} and~\eqref{EqSecondOrder} are equivalent to Eqs.~\eqref{eq:theo1} and~\eqref{eq:theo2} of the main text. They are used to generate the theoretical predictions shown in 
Fig.~\ref{fig:theoryVSexpVSsimu} and Fig.~\ref{fig:tl1}. \\

%%%%%%%%%%%%%%%%%%%%%%%%%%%%%%%%%%%%%%%%%%%%%%%%%%%%%%%%%%%%%%%%%%%%
%%%%%%%%%%%%%%%%%%%%%%%%%%% ANNEXE D %%%%%%%%%%%%%%%%%%%%%%%%%%%%%%%
%%%%%%%%%%%%%%%%%%%%%%%%%%%%%%%%%%%%%%%%%%%%%%%%%%%%%%%%%%%%%%%%%%%%
\section{RME characterization of different graded-index fibers}
\label{Appendix:GRIN}
In this Appendix, we report the measurements for different fiber segments of the same length ($L=24.5$~cm), and with advertised properties similar to those of the fiber used in the main text. Specifically, we use samples from a 
Thorlabs 50-micron core OM2 graded-index fiber (GIF50C, NA = 0.2).

\begin{figure}[t]
\includegraphics[width=0.49\textwidth]{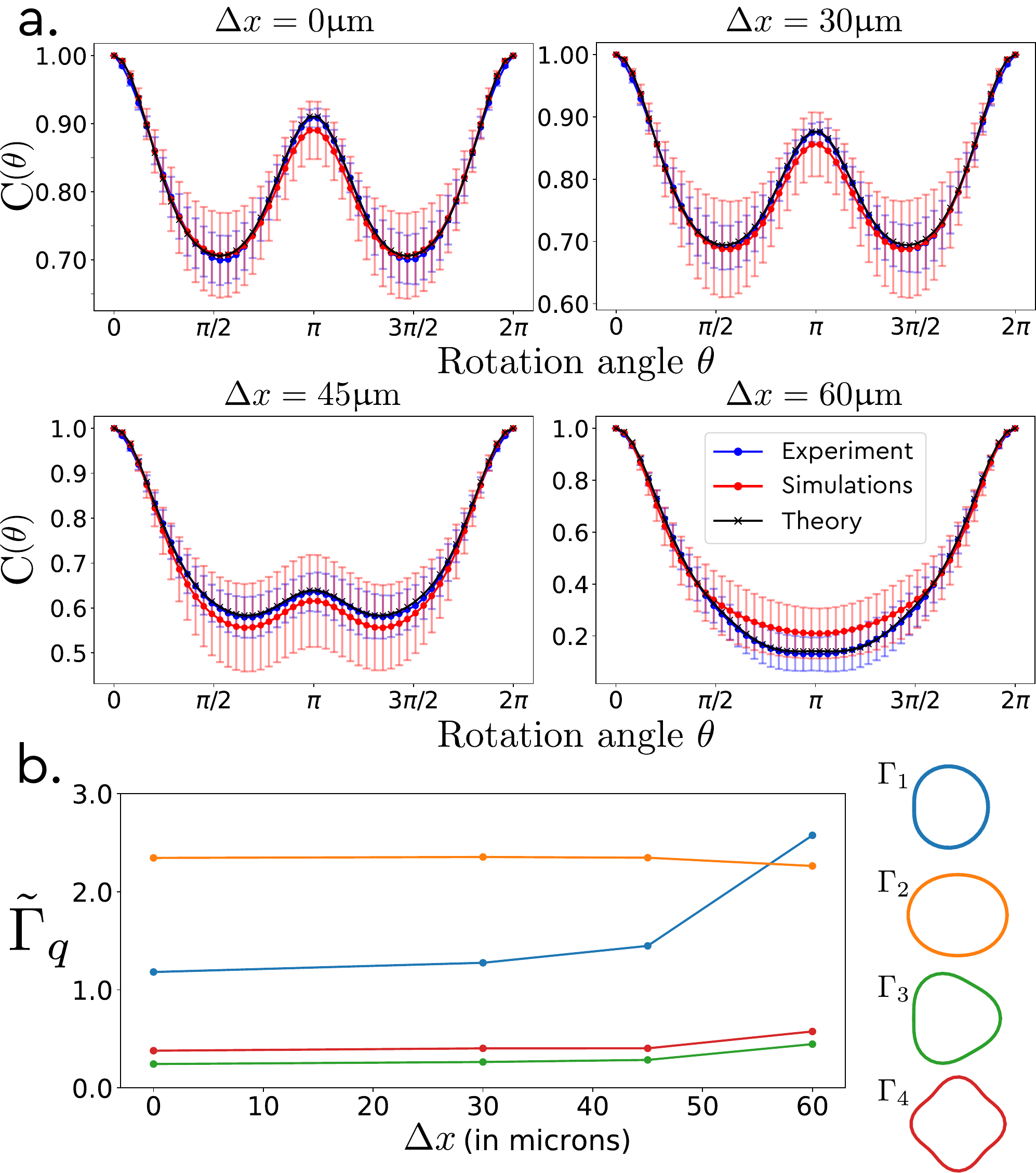}
\caption{
\textbf{RME correlation results for a batch of GIF50C.}
    (a) Angular correlation function of the RME, as defined in Eq.~\eqref{eq:corr_function},
    for various levels of deformation $\Delta x$.
    Experimental data (blue) are compared to
    the theoretical prediction based on Eqs.~\eqref{eq:theo1} and~\eqref{eq:theo2} (black),
    and to simulation results obtained with the same parameters as those used in the theoretical model (red).
    (b)  Values of the normalized deformation parameters 
    $\tilde{\Gamma}_q = k l_z \sigma_g(r=a)  \Gamma_q$.
    The values of $\Gamma_q$ 
    are found by fitting
    the theoretical model [Eq.~\eqref{eq:theo1}] to the experimental data as a function of the deformation.
    In the inset, we show the symmetry corresponding to the perturbation associated with each value of $q$.
}
\label{fig:tl1}
\end{figure}

Results for different fiber segments of the same spool are reproducible. 
We present typical results for one sample in Fig.~\ref{fig:tl1}.
We observe different contributions of the $\Gamma_q$ terms as the ones reported in Fig.~\ref{fig:theoryVSexpVSsimu}, where a  Prysmian BendBright OM4 fiber was used~\cite{bendbright_OM4}.
In particular, $\Gamma_4$ is much smaller,
leading to the absence of observed local maxima
of the correlation at $\pi/2$ and $3\pi/2$.

%%%%%%%%%%%%%%%%%%%%%%%%%%%%%%%%%%%%%%%%%%%%%%%%%%%%%%%%%%%%%%%%%%%%
%%%%%%%%%%%%%%%%%%%%%%%%%%% ANNEXE E %%%%%%%%%%%%%%%%%%%%%%%%%%%%%%%
%%%%%%%%%%%%%%%%%%%%%%%%%%%%%%%%%%%%%%%%%%%%%%%%%%%%%%%%%%%%%%%%%%%%
\section{Properties of the RME channels}

\subsection{RME Operator for one angle value}
\label{Appendix:OperatorOneAngle}

\begin{figure}[t]
\includegraphics[width=0.47\textwidth]{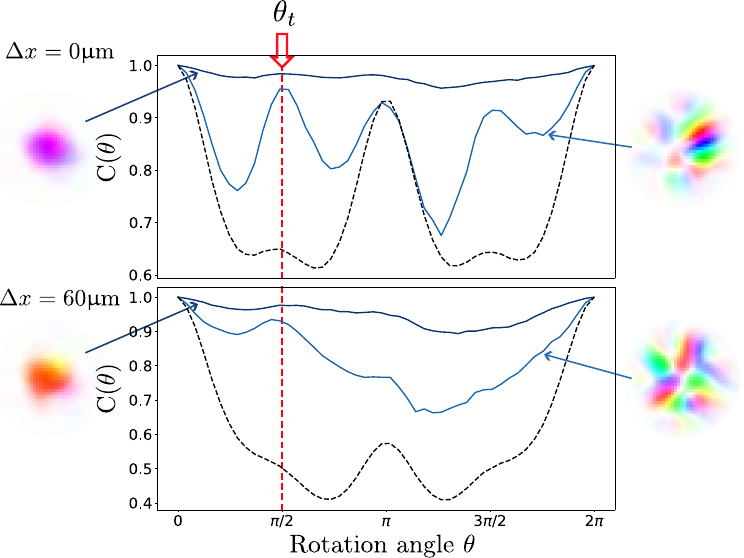}
\caption{
\textbf{Tailoring the rotational memory effect.}
The angular correlation function $C(\theta)$ is constructed using experimentally measured input channels with 
improved RME range, for two values of the deformation ($\Delta x=0\,\mu$m and $\Delta x=60\,\mu$m). The results for the first two  singular vectors of the operator defined in Eq.~\eqref{eq:RME_sum_op}  are compared with the average results for random input profiles (dashed line). 
The insets show the output spatial transverse profiles of the corresponding singular vectors. 
}
\label{fig:tailoring_one_angle}
\end{figure}

We consider here operator $\mathbf{O}(\theta_t)$ defined in Eq.~\eqref{eq:RME_op}, which represents the upper part of the
correlation function~\eqref{eq:corr_function}.
Computing the singular values of this operator
enables the identification of input wavefronts that maximize the
angular correlation for a specific value $\theta_t$ of $\theta$.
We present in Fig.~\ref{fig:tailoring_one_angle} the resulting correlation $C(\theta)$ of
the first two singular vectors for $\theta_t=\pi/2$,
in the cases of no deformation and strong deformations ($\Delta x = 60$\textmu m),
along with the corresponding output field profiles.
As with the results presented for the sum operator
in Fig.~\ref{fig:tailoring},
the first singular vector of the operator~\eqref{eq:RME_op} closely resembles the fundamental mode for any $\Delta x$.
The second singular vector exhibits higher spatial frequencies and
achieves a maximum in the correlation at the target angular value $\theta_t$.
However, although the correlation curve is consistently higher than the average correlation,
it displays significant fluctuations over the $2\pi$ range.

\subsection{Autocorrelation of the RME channels}
\label{Appendix:AutoCorrelation}

As stated in the main text, for efficient information retrieval from the hidden side of a complex medium, 
the output pattern used should possess both a substantial memory effect range and a narrow autocorrelation function.
This implies that the correlator $C(\theta)$ defined in Eq.~\eqref{eq:corr_function} displays a broad width, 
and that the angular autocorrelation
\be
C_0(\theta) = \frac{
                \vert\bra{\psi} 
                    \mathbf{R}(\theta)
                \ket{\psi}\vert
                }
                {
                    \ps{\psi}{\psi}
                }
\ee
 of the transmitted field $\ket{\psi}=\mathbf{T}\ket{\psi_\text{in}}$ is sharply peaked.

\begin{figure}[t]
\includegraphics[width=0.49\textwidth]{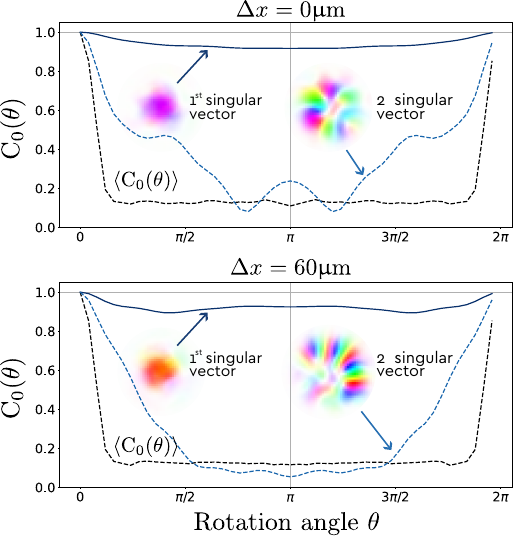}
\caption{
\textbf{Autocorrelation of the RME channels.}
Experimental rotational autocorrelation of the output field when injecting the wavefront associated with
the first (solid dark blue line) 
and  second  (dashed light blue line) highest singular modes,
without and with external deformations.
For comparison, we also show the average autocorrelation 
of the output field for random input wavefront (dashed black line), 
that represents the optimal autocorrelation.
}
\label{fig:autocorr}
\end{figure}

We present in Fig.~\ref{fig:autocorr} the autocorrelation of the first two RME eigenchannels of the operator $\mathbf{O}_\text{sum}$ defined in Eq.\eqref{eq:RME_sum_op}, under different conditions 
— absence of external deformations and presence of strong deformations ($\Delta x=60\,\mu$m). 
As anticipated in the main text, the first singular mode, being less susceptible to perturbations, 
exhibits a rotational symmetry with a strong correlation across the entire angular range, 
even under significant deformations. 
This indicates that this mode, despite its wide RME range, does not meet the criteria defined in the main text to be a viable candidate for image or information retrieval.
However, the second singular mode does exhibit a pronounced RME peak around $\theta = 0$. 
The combination of this property with the wide angular range of the RME makes this mode an excellent candidate for effective use in discerning the angular properties of an object or signal obscured at the fiber distal end.

\end{document}